\documentclass[aps,prl,amsfonts,amsmath,amssymb,nofootinbib,twocolumn,superscriptaddress]{revtex4-2}

\usepackage{braket}
\usepackage{graphics}
\usepackage{hyperref}
\usepackage{epsfig}
\usepackage{color}
\usepackage{verbatim}

\usepackage{ulem}
\usepackage{multirow}
\usepackage{bm}

\usepackage{pifont}
\usepackage{amssymb}
\usepackage{amsmath}
\usepackage[version=4]{mhchem}

\begin{document}

\title{Observation of Room-temperature Charge Density Wave Correlations via Coherent Phonon Spectroscopy in Sn-doped Kagome Superconductor CsV$_3$Sb$_5$}

\author{Qinwen Deng}
\affiliation{Department of Physics and Astronomy, University of Pennsylvania, Philadelphia, Pennsylvania 19104, U.S.A.}
\author{Andrea Capa Salinas}
\affiliation{Materials Department, University of California Santa Barbara, Santa Barbara, California
93106, U.S.A.}
\author{Suchismita Sarker}
\affiliation{Cornell High Energy Synchrotron Source, Cornell University, Ithaca, New York 14853, U.S.A.}
\author{Leon Balents}
\affiliation{Kavli Institute for Theoretical Physics, University of California Santa Barbara, Santa Barbara, California
93106, U.S.A.}
\author{Stephen D. Wilson}
\affiliation{Materials Department, University of California Santa Barbara, Santa Barbara, California
93106, U.S.A.}
\author{Liang Wu}
\email{liangwu@sas.upenn.edu}
\affiliation{Department of Physics and Astronomy, University of Pennsylvania, Philadelphia, Pennsylvania 19104, U.S.A.}

\date{\today}

\begin{abstract}

In this work, we perform ultrafast time-resolved reflectivity measurements to track the evolution of charge density wave (CDW) correlations in Sn-doped Kagome superconductor CsV$_3$Sb$_{5-x}$Sn$_x$. By extracting the coherent phonon spectrum, we evidence robust signatures of CDW correlations at temperature and doping ranges far beyond the phase boundary of long-range CDW order. Remarkably, we unveil short-range CDW correlations survive up to room temperature in $x = 0.32$ Sn-doped CsV$_3$Sb$_5$, supported by synchrotron X-ray diffraction measurements. We point out the introduction of quenched disorder by Sn doping can pin the CDW and form static short-range CDW, which can explain the observed persistent CDW signatures. Our results thus corroborate the ubiquity and robustness of CDW correlations in Sn-doped CsV$_3$Sb$_5$ and provide new insights on the role of disorders on the CDW correlations in AV$_3$Sb$_5$ family. 
\end{abstract}

\maketitle 

\section*{I. Introduction}
The newly discovered Kagome metals $A$V$_3$Sb$_5$ ($A$ = K, Rb, Cs) has attracted tremendous research interest by exhibiting an unconventional electronic landscape, including charge density wave (CDW) and superconductivity\cite{ortiz2019new, ortiz2020cs, ortiz2021superconductivity, yin2021superconductivity}. The CDW is fundamental to the exotic phenomena in these compounds, as it sets the stage for a cascade of intertwined symmetry breaking orders, including possible time-reversal symmetry breaking\cite{mielke2022time, jiang2021unconventional, xing2024optical, guo2022switchable, xu2022three, park2021electronic, feng2021chiral, denner2021analysis, lin2021complex, guo2024correlated, christensen2022loop}, electronic nematicity\cite{nie2022charge, zhao2021cascade, li2023unidirectional, xiang2021twofold, asaba2024evidence, liu2024absence}, and pair density waves\cite{chen2021roton}. Therefore, understanding the evolution of CDW under external perturbations is of paramount significance to clarify the interactions essential for stabilizing this plethora of electronic orders. 

\begin{figure*}
    \centering
    \includegraphics[width=\textwidth]{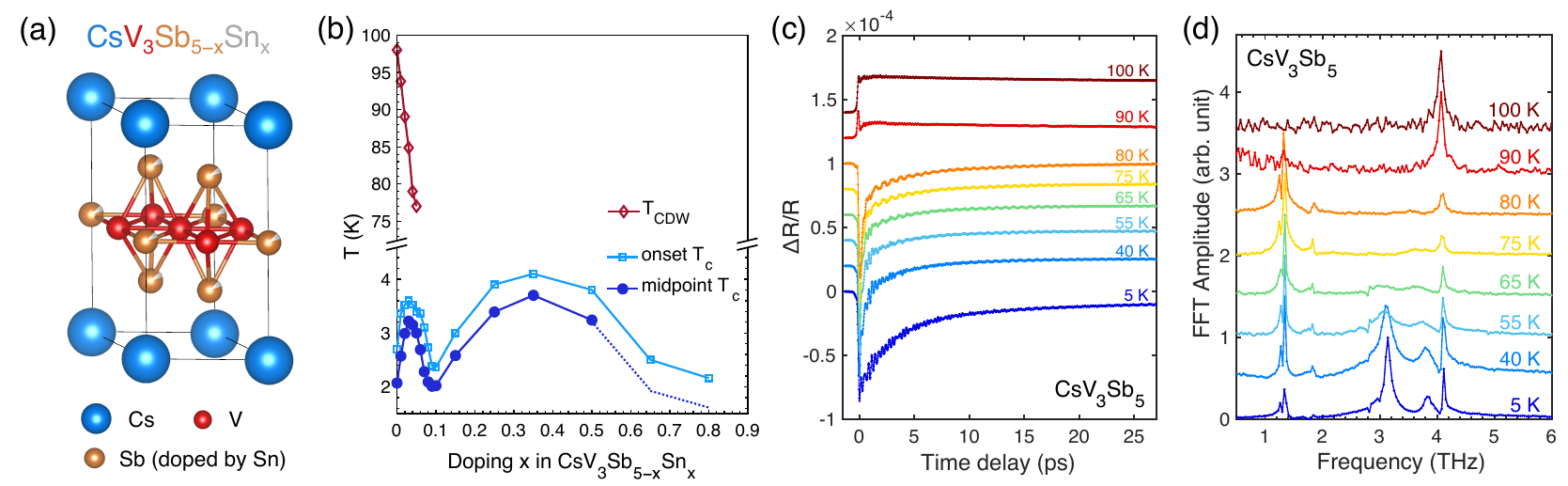}
    \caption{(a) Crystal structure of CsV$_3$Sb$_{5-x}$Sn$_x$. (b) Temperature-doping phase diagram for CsV$_3$Sb$_{5-x}$Sn$_x$. (c) $\Delta$R/R vs. temperature in undoped CsV$_3$Sb$_5$\cite{deng2025coherent}. (d) Temperature dependence of coherent phonon spectra in undoped CsV$_3$Sb$_5$\cite{deng2025coherent}. 
    }
    \label{fig1}
\end{figure*}

Chemical doping provides a powerful route to tune the CDW order in $A$V$_3$Sb$_5$ and study its interplay with other correlated states\cite{oey2022fermi, kautzsch2023incommensurate, zhong2023nodeless, liu2023doping, yang2022titanium, xiao2023evolution, liu2022evolution, ding2022effect, li2022tuning, huai2024two, huai2025electronic}. In CsV$_3$Sb$_{5-x}$Sn$_x$ focused by this study, Sn atoms substitute on Sb sites (Fig. \ref{fig1}a). Increasing $x$ rapidly suppresses the long-range CDW order, which disappears in thermodynamic measurements above $x = 0.06$ (Fig. \ref{fig1}b)\cite{oey2022fermi}. Concurrently, the superconducting transition temperature $T_c$ initially rises with light doping, indicating direct competition between superconductivity and CDW as also reported in pressure\cite{yu2021unusual, chen2021double} and strain\cite{qian2021revealing}-dependent studies. When further increasing Sn doping, $T_c$ shows a double-dome feature, with maximums at $x \simeq 0.03$ and $0.35$ and a minimum at $x \simeq 0.09$. Moreover, quasi-1D incommensurate short-range charge correlations emerge in $x = 0.15$ sample without long-range CDW\cite{kautzsch2023incommensurate}. However, in Sn-doped CsV$_3$Sb$_5$ there still lacks more systematic and detailed studies on how CDW correlations evolve across an extended temperature and doping range, and the possible role of dopants on the CDW. 

Beyond the widely-discussed hole doping effect, Sn doping into CsV$_3$Sb$_5$ also introduces quenched disorder which has profound impacts on CDW\cite{harris1974effect, imry1975random, brock1994detailed, vojta2013phases}, such as local pinning effects\cite{kivelson2003detect, nie2015fluctuating, arguello2014visualizing, liu2021thermal, yue2020distinction, okamoto2015experimental}. Here, we use ultrafast time-resolved reflectivity (TR-reflectivity) to track the evolution of CDW correlations in CsV$_3$Sb$_{5-x}$Sn$_x$. Coherent phonon spectra extracted from TR-reflectivity time traces reveal robust and ubiquitous signatures of CDW correlations well beyond the disappearance of long-range CDW. Notably, we detect evidence of short-range CDW persisting to room temperature in the $x = 0.32$ compound, supported by synchrotron X-ray diffraction. We point out the introduction of quenched disorder by Sn doping can pin the CDW and form static short-range CDW, explaining the observed persistence of CDW correlations across a broad phase space. 

%\red{In Fig.a, make the crystal smaller. Done. Perhaps you can put those atom lables on top of the crystal structure so that you can make c,d wider (larger). Done. I think you also need to make those labels such as the temperature larger. Done. } 

\begin{figure*}
    \centering
    \includegraphics[width=\textwidth]{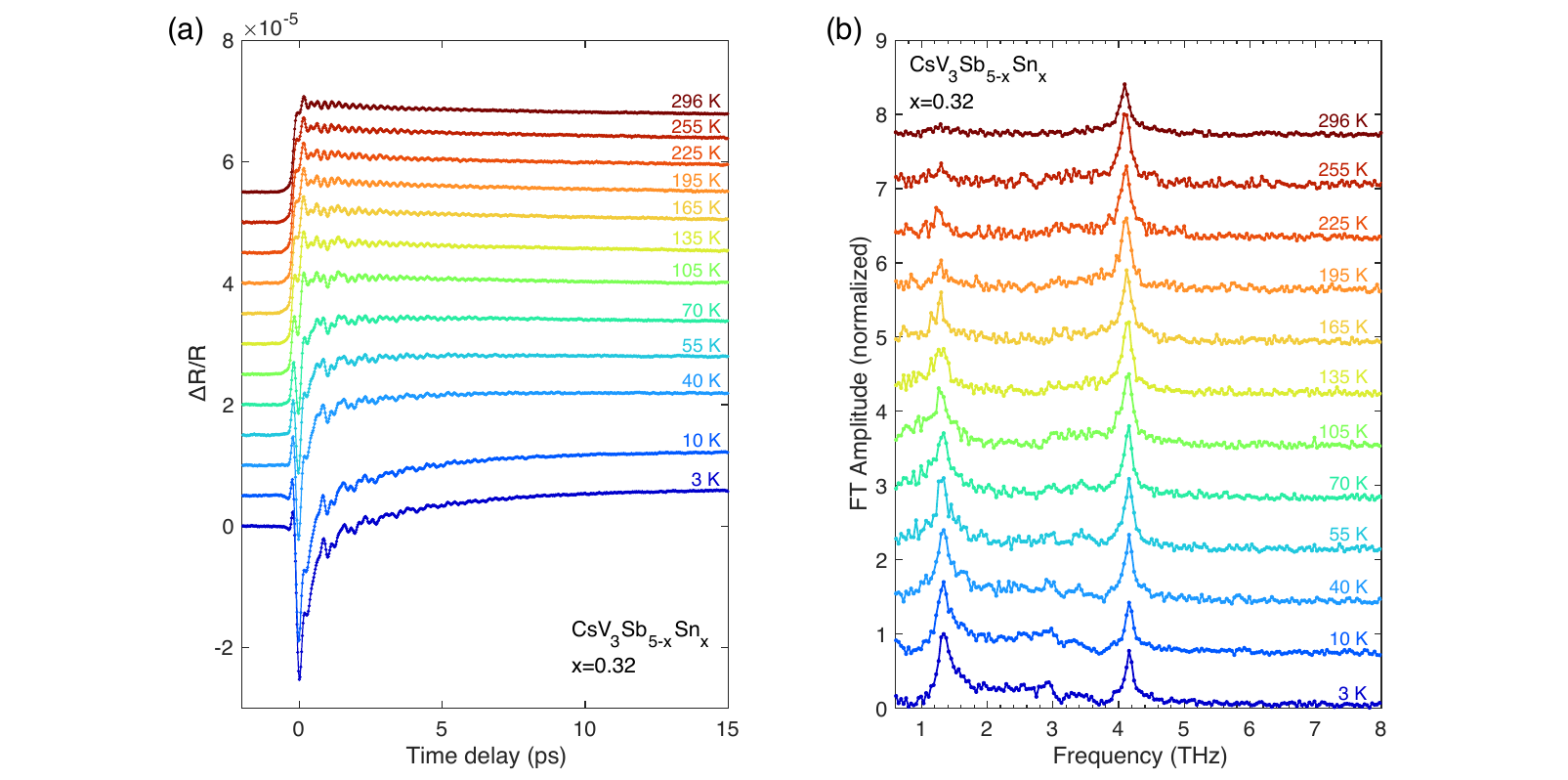}
    \caption{\textbf{Evolution of the coherent phonon spectrum in CsV$_3$Sb$_{5-x}$Sn$_x$ with $x = 0.32$ with temperature. }(a) TR-reflectivity curves $\Delta$R/R in the temperature range of 3 K – 296 K. (b) Temperature dependence of  Coherent phonon spectra. The curves are offset for clarity. }
    \label{fig2}
\end{figure*}

% Single crystals of CsV$_3$Sb$_{5-x}$Sn$_x$ with different doping level $x$ were synthesized using the self-flux method\cite{ortiz2019new, ortiz2021superconductivity}. We perform TR-reflectivity measurements on freshly-cleaved (001) surface of these Sn-doped CsV$_3$Sb$_5$ single crystals, using pump wavelength of 1560 nm and probe wavelength of 780 nm, with repetition rate of 80 MHz and pulse duration of 100 fs. Both beams are at normal incidence with $\sim$10 $\mu$m spot diameter and fluences less than 10 $\mu$J/cm$^2$ to minimize heating. %The pump beam intensity is modulated at 84 kHz using a photo-elastic modulator and a pair of linear polarizers. 
% Temperature-dependent, synchrotron X-ray diffraction measurements were performed at the Cornell High Energy Synchrotron Source (CHESS) on the QM2 (ID4B) beamline. Diffraction data was acquired in transmission mode using incident $\lambda$ = 0.47686 Å (26.0 keV) and collected using a 6-megapixel detector with 0.1 s exposure time. The sample was cooled using a nitrogen/helium-switchable cryostream. Data reduction was performed with the NXRefine GUI software. 

CsV$_3$Sb$_{5-x}$Sn$_x$ single crystals with different doping level $x$ were synthesized using the self-flux method\cite{ortiz2019new, ortiz2021superconductivity}. We perform TR-reflectivity measurements on freshly-cleaved (001) surface of these Sn-doped CsV$_3$Sb$_5$ single crystals, using 1560 nm pump and 780 nm probe pulses with repetition rate of 80 MHz and pulse duration of 100 fs\cite{deng2025coherent, deng2025revealing}. Both beams are at normal incidence with $\sim$10 $\mu$m spot diameter and fluences below 10 $\mu$J/cm$^2$ to minimize heating. Synchrotron X-ray diffraction measurements were performed at the Cornell High Energy Synchrotron Source (CHESS) on the QM2 (ID4B) beamline in transmission geometry with incident $\lambda$ = 0.47686 Å. Diffraction data was collected using a 6-megapixel detector with 0.1 s exposure time. The sample was cooled using a nitrogen/helium-switchable cryostream. Data reduction was performed with the NXRefine GUI software\cite{NXRefine}. 

%In the pristine phase, CsV$_3$Sb$_{5-x}$Sn$_x$ with $0 \leq x \leq 1 $ host P6/mmm space group with a V Kagome network and the Sn atoms are doped into Sb sites, introducing hole doping. 

\begin{figure}
    \centering
    \includegraphics[width=8.6cm]{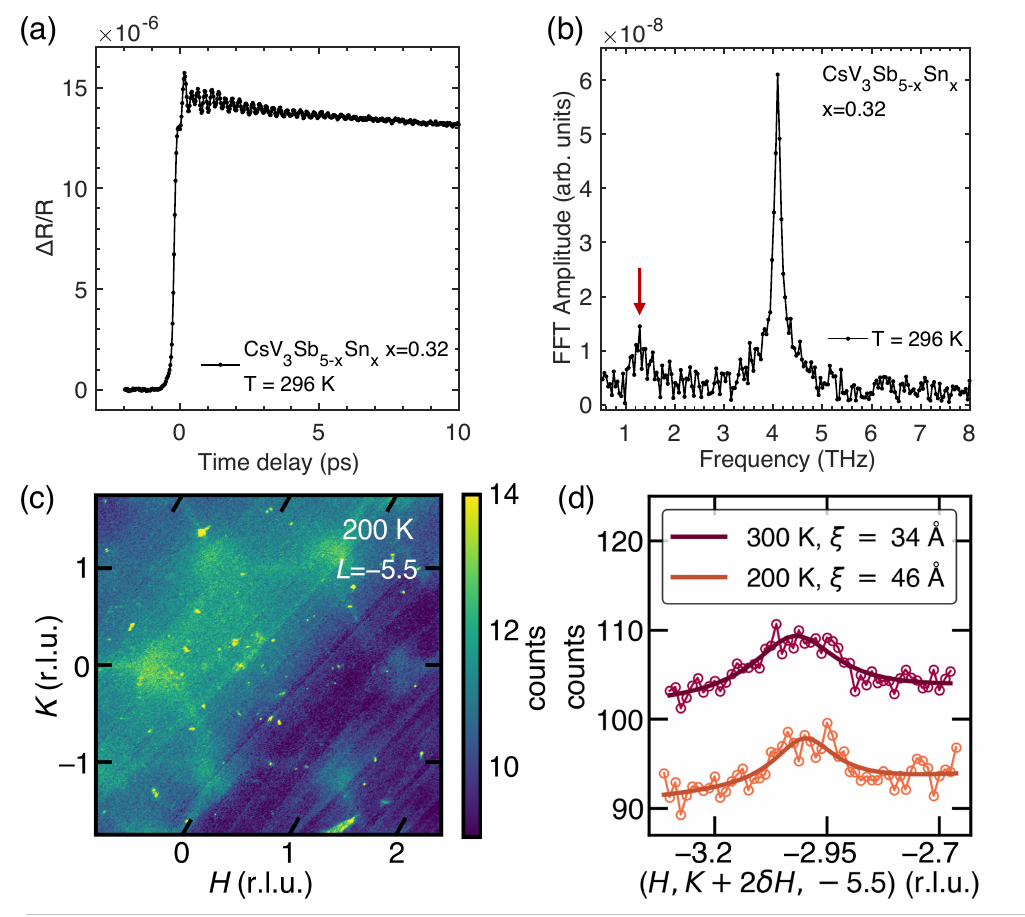}
    \caption{\textbf{Room temperature short-range charge orders in CsV$_3$Sb$_{5-x}$Sn$_x$ with $x = 0.32$. }(a) Zoom-in of $\Delta$R/R at $T$ = 296 K. (b) Coherent phonon spectrum at $T$ = 296 K. The red arrow marks the 1.3 THz mode. (c) X-ray $(H, K)$-map at $L = -5.5$ for CsV$_3$Sb$_{5-x}$Sn$_x$, $x = 0.32$ at 200 K. (d) Representative linecuts perpendicular to the stripe direction (along [-1, 2, 0]) on the peak (-3, 1.6, -5.5) at 200 K and 300 K with correlation lengths $\xi$. Fits were performed using Lorentzian functions with a linear background.}
    \label{fig3}
\end{figure}

\section{II. EXPERIMENT RESULTS ON S\MakeLowercase{n}-doped C\MakeLowercase{s}V$_3$S\MakeLowercase{b}$_5$}
%\section{II. EXPERIMENT RESULTS ON Sn-doped CsV$_3$Sb$_5$}
TR-reflectivity sensitively probes CDW, since modifications of the electronic structure and phonon spectrum due to CDW can manifest in transient reflectivity changes. Fig. \ref{fig1}c shows the transient reflectivity change ($\Delta$R/R) versus time delay in undoped CsV$_3$Sb$_5$\cite{deng2025coherent}. The sign of initial reflectivity change $\Delta$R/R(t = 0) flips across $\approx$ 90 K, consistent with reported $T$$_{\text{CDW}}$ = 94 K with the small offset attributable to local laser heating. This sign change likely reflects band renormalization and partial bandgap opening at the CDW transition\cite{zhou2021origin, nakayama2021multiple, uykur2021low}. The corresponding coherent phonon spectra (Fig. \ref{fig1}d)\cite{deng2025coherent} reveal that above $T$$_{\text{CDW}}$ only the 4.1 THz main lattice mode remains, whereas below $T$$_{\text{CDW}}$ two additional intense modes at 1.3 and 3.1 THz appear at 5 K, both showing dual peaks indicating coexistence of Star-of-David and inverse Star-of-David distortions with broken $C_6$ symmetry\cite{deng2025coherent}. %We also observe weaker modes at 3.86 and 1.84 THz. 
The 1.3 THz mode persists up to $T$$_{\text{CDW}}$, indicating it's CDW-induced. Its minimal frequency softening upon warming agrees with a CDW-induced zone-folded phonon mode\cite{joshi2019short}. %The 3.1 THz modes weaken and broaden during warm-up and disappear at $\approx$ 65 K, also observed by Raman measurements\cite{liu2022observation, he2024anharmonic}. 
% , agreeing with a 4.10 THz $A_{1g}$ mode in the pristine phase of CsV$_3$Sb$_5$
We focus on the $\Delta$R/R sign reversal and the 1.3 THz mode as key indicators of CDW.

We next perform TR-reflectivity measurements on CsV$_3$Sb$_{5-x}$Sn$_x$ with $x = 0.32$. This sample is well beyond the boundary of long-range CDW in the temperature-doping phase diagram\cite{oey2022fermi}, where thermodynamic measurements show no long-range CDW. Yet unexpectedly, we observe clear signatures of CDW. Fig. \ref{fig2}a shows the evolution of $\Delta$R/R with temperature. At $T$ = 3 K, $\Delta$R/R initially drops to negative values at t = 0, then rapidly relaxes and turns to positive. With warming, the initial negative drop gradually diminishes and the whole $\Delta$R/R curve becomes positive near 70 K, though a small drop after the initial rise persists up to $\sim$ 165 K. %\red{I probably asked you this before. What drop are you talking about between 70K and 165K?). Done. } 
This sign switch is much slower than in undoped CsV$_3$Sb$_5$ where $\Delta$R/R(t = 0) sign flips abruptly within 5 K at $T$$_{\text{CDW}}$\cite{ratcliff2021coherent}. We then subtract the decaying background and extract the coherent phonon spectrum by performing Fourier transform of the oscillation parts (Fig. \ref{fig2}b). In AV$_3$Sb$_5$ family, coherent phonons are generated in pump-probe experiments via displacive excitation of coherent phonons (DECP)\cite{deng2025coherent, ratcliff2021coherent, deng2025revealing} where only fully symmetric modes can be observed\cite{cheng1991mechanism, zeiger1992theory}. DFT calculation shows only one fully-symmetric $A_{1g}$ mode at 4.10 THz in the pristine phase of CsV$_3$Sb$_5$\cite{deng2025coherent} which is the only observable mode without CDW. However, at 3 K, besides this main lattice mode near 4.1 THz, we observe an additional prominent phonon mode near 1.3 THz (Fig. \ref{fig2}b). This indicates the existence of CDW correlation which induces this mode. This 1.3 THz mode exhibits a much broader linewidth than in undoped CsV$_3$Sb$_5$ and decays in $\Delta$R/R time trace within $\sim$ 7 ps, contrary to over 60 ps in undoped CsV$_3$Sb$_5$ with long-range CDW\cite{deng2025coherent}. Weak and broad peak-like features also emerge at $\sim$ 3 THz but vanish above 55 K. We do not detect other modes at frequencies higher than the 4.1 THz main lattice mode. Remarkably, although relatively weaker during warming, this 1.3 THz mode persists up to room temperature $T$ = 296 K, which is more than triple the long-range CDW transition temperature $T$$_{\text{CDW}}$ = 94 K in undoped CsV$_3$Sb$_5$\cite{uykur2021low, ortiz2020cs}. This mode also shows minimal frequency softening from 3 K to 296 K, matching with a zone-folded ordinary phonon induced by CDW correlations.

\begin{figure*}
    \centering
    \includegraphics[width=\textwidth]{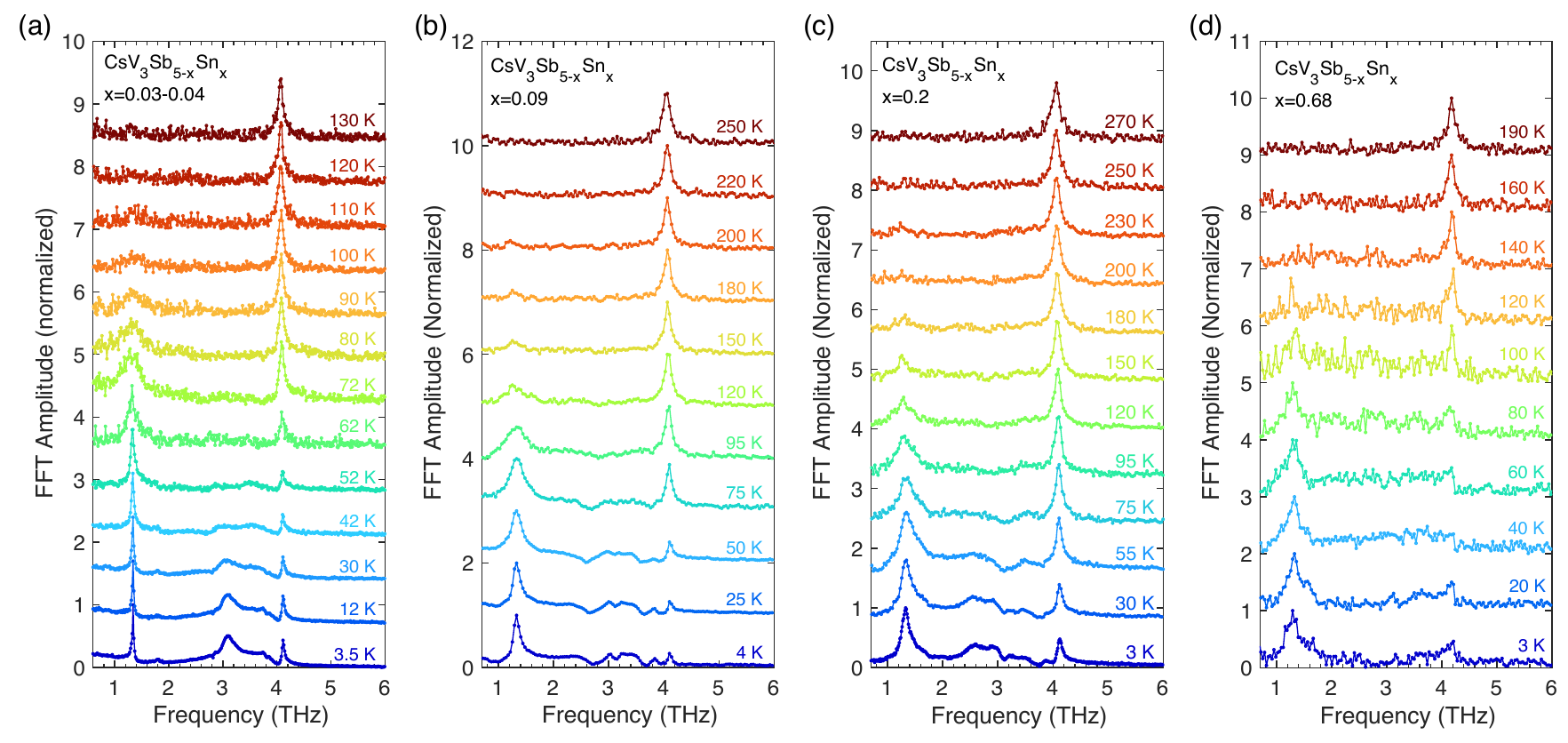}
    \caption{\textbf{Coherent phonon spectra among different doping levels of CsV$_3$Sb$_{5-x}$Sn$_x$} with (a) $x = 0.03-0.04$ (b) $x = 0.09$ (c) $x = 0.2$ and (d) $x = 0.68$ vs. temperature. The curves are offset for clarity. }
    \label{fig4}
\end{figure*}

Fig. \ref{fig3}a, b show the zoom-in of $\Delta$R/R trace and corresponding coherent phonon spectrum of the $x = 0.32$ Sn-doped CsV$_3$Sb$_5$ at $T$ = 296 K respectively. The first 2 ps of the time trace reveal more than one oscillation component. Fig. \ref{fig3}b confirms the presence of the 1.3 THz mode above the noise level. Synchrotron X-ray diffraction further evidences short-range charge orders up to 300 K. Fig. \ref{fig3}c displays the $H-K$ map of the scattering data on the $L = -5.5$ plane at $T$ = 200 K on this CsV$_3$Sb$_{5-x}$Sn$_x$ with $x = 0.32$ sample, where we reveal quasi-1D patterns of CDW correlations along all three in-plane axes that are 120$^\circ$ to each other. These charge correlations show an incommensurate short-ranged state, as indicated by a representative linecut perpendicular to the stripe direction (along [-1, 2, 0]) on the peak (-3, 1.6, -5.5) (Fig. \ref{fig3}d) with correlation length $\xi = 46$ Å at 200 K. At 300 K, the CDW correlation is still visible with $\xi = 34$ Å. These behaviors indicate three-state short-range domains that are 120$^\circ$ rotated in real space, where charge orders form along a specific in-plane axis in each domain thereby breaking $C_6$ rotational symmetry down to $C_2$. These features resemble the quasi-1D charge scattering along three in-plane axes reported in $x = 0.15$ Sn-doped CsV$_3$Sb$_5$\cite{kautzsch2023incommensurate} but no charge correlations is seen at 260 K therein, whereas here short-range charge correlations persist up to room temperature in $x = 0.32$ Sn-doped compound. The observed shorter lifetime of the 1.3 THz mode in this sample is also consistent with this significantly reduced correlation length compared to undoped CsV$_3$Sb$_5$\cite{kautzsch2023structural}. In undoped CsV$_3$Sb$_5$, the 1.3 THz mode comes from folding an L-point phonon to $\Gamma$ by CDW, and DFT shows this phonon has minimal dispersion between A-L\cite{ratcliff2021coherent}. Thus, any CDW order with $q_z = 0.5$ r.l.u. and some uniaxial in-plane modulation would activate this mode and cause it to appear in the coherent phonon spectrum. This explains our observed mode near 1.3 THz in the $x = 0.32$ Sn-doped compound.  

\begin{figure}
    \centering
    \includegraphics[width=8.6cm]{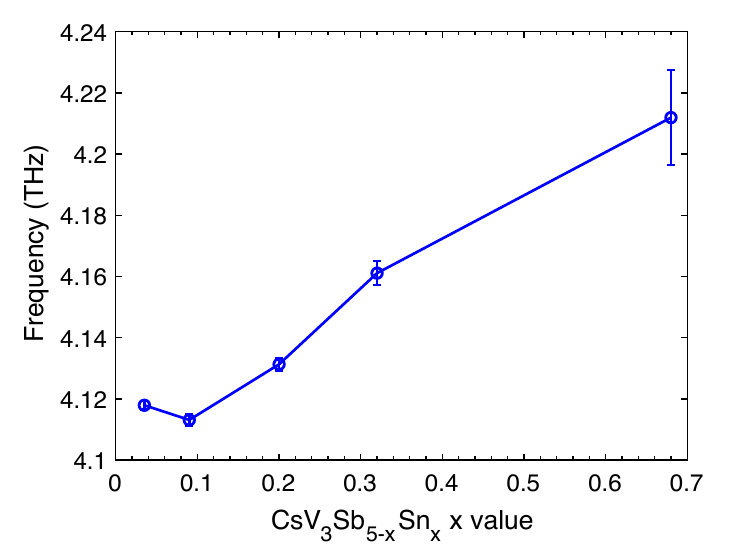}
    \caption{The main lattice mode frequency as a function of the doping level $x$ in CsV$_3$Sb$_{5-x}$Sn$_x$.}
    \label{fig5}
\end{figure}

We extend TR-reflectivity and coherent phonon spectra studies across a wider doping range and track the evolution of CDW correlations (Fig. \ref{fig4}). Among all four doping levels, only CsV$_3$Sb$_{5-x}$Sn$_x$ with x = 0.03-0.04 shows long-range CDW order\cite{oey2022fermi}. The bulk $T$$_{\text{CDW}}$ in CsV$_3$Sb$_{5-x}$Sn$_x$ is 85 K with $x = 0.03$ and 79 K with $x = 0.04$\cite{oey2022fermi}. In the $x = 0.03-0.04$ sample (Fig. \ref{fig4}a), we observe a single 1.3 THz mode, 3.1 THz dual modes, and weaker peaks near 1.8 and 3.8 THz at $T$ = 3.5 K, consistent with the ISD + ISD with interlayer $\pi$-phase shift\cite{deng2025coherent}. Intriguingly, the 1.3 THz mode persists above bulk $T$$_{\text{CDW}}$ unlike in undoped CsV$_3$Sb$_5$. Despite substantially weakened and damped, this 1.3 THz mode survives up to 110 K, indicating short-range CDW correlations above bulk $T$$_{\text{CDW}}$. In higher dopings (CsV$_3$Sb$_{5-x}$Sn$_x$ with $x = 0.09$, 0.2 and 0.68) lacking long-range CDW according to thermodynamic measurements (Fig. \ref{fig4}b-d), besides the 4.1 THz main lattice mode, the dominant low-$T$ feature is again the CDW correlation-induced 1.3 THz mode. Their linewidth is much larger compared to samples with long-range CDW, also suggesting reduced correlation lengths. The main lattice mode frequency shows an increasing trend as $x$ increases (Fig. \ref{fig5}). Since the 4.1 THz main lattice mode is linked to the motion of the Sb atoms\cite{ratcliff2021coherent}, when the atomic mass decreases when substituting Sb by Sn, the vibration frequency is expected to increase, matching with the observed trend. Additional weak and broad peaks near 3-4 THz appear in $x = 0.09$ and $x = 0.2$ samples but vanish by 75-95 K, whereas there is no well-defined modes between 3-4 THz in the $x = 0.68$ sample. With warming, the 1.3 THz peak weakens and broadens, disappearing at $T$ = 220 K ($x = 0.09$), 250 K ($x = 0.2$), and 140 K ($x = 0.68$). %\red{Emphasize here that when x=0.09 where the TC is mimimum, the short-range CDW order is not the strongest. John Harter's paper is not right. Done. } 
This indicates CDW correlations in these compounds without long-range CDW survive at temperatures much higher than $T$$_{\text{CDW}}$ in undoped CsV$_3$Sb$_5$. Notably, in the lowest $T_c$ sample ($x = 0.09$), the onset temperature of CDW correlation is not maximal, seemingly contrary to the competition scenario. In summary, our coherent phonon spectrum study discovers ubiquitous and robust CDW correlations in Sn-doped CsV$_3$Sb$_5$ persisting far beyond the temperature and doping limit of long-range CDW. 

\section{III. Discussions}

It is, therefore, necessary to discuss the important role of Sn doping on the emergence of CDW correlations beyond the temperature and doping range of long-range CDW. Apart from hole doping, Sn dopants into CsV$_3$Sb$_5$ also introduce quenched disorder into the system. In mean-field theory, quenched disorders may have different forms of coupling to the order parameter, manifesting in the following Landau-Ginzburg-Wilson (LGW) free-energy functional\cite{harris1974effect, imry1975random, brock1994detailed, vojta2013phases}: 
\begin{align}
\mathcal{F}[\psi(\vec{x})] 
&= \int d^d x \big\{-h(\vec{x})\psi(\vec{x}) 
    + [r + \delta r(\vec{x})]\psi^2(\vec{x}) \notag \\
&\quad + (\nabla\psi(\vec{x}))^2 
    + u\psi^4(\vec{x}) + \ldots \big\}
\end{align}
Here, $h(\vec{x})$ is associated with the random-field disorder, and $\delta r(\vec{x})$ is associated with random-mass disorder. The random-field disorder pins the phase of CDW, while random-mass disorder changes the local critical temperature, i.e. the disorder can change the local tendency towards ordering. Although the average effect of disorder may suppress long-range CDW orders, the spatial variation caused by disorder may locally prefer CDW as lowering of CDW free energy can happen at local points, thus allowing local short-range CDW orders. Indeed, static short-range CDW orders can survive well above $T$$_{\text{CDW}}$ in the vicinity of defects\cite{oh2020defect, chatterjee2015emergence, arguello2014visualizing} %NC volume 6, Article number: 6313 (2015), PRB 89, 235115 (2014).] 
e.g. up to 3$T$$_{\text{CDW}}$ in 2H-NbSe$_2$\cite{arguello2014visualizing}, and above the critical doping level $x_C$ where $T$$_{\text{CDW}}$ drops to 0\cite{chatterjee2015emergence}. %[NC volume 6, Article number: 6313 (2015)]

A recent TR-reflectivity study proposes CDW fluctuations to explain the observation of CDW-induced phonon modes beyond the long-range CDW phase boundary in CsV$_3$Sb$_{5-x}$Sn$_x$\cite{kongruengkit2026persistence}. Such CDW fluctuations (dynamic charge order) are typically probed by resonant inelastic X-ray scattering\cite{comin2016resonant}. However, our X-ray scattering uses $\lambda$ = 0.47686 Å (26 keV) to avoid resonant absorption edges, and the data is energy integrated so that we cannot confirm dynamic charge order that is manifested by inelastic scattering. Alternatively, quenched disorder from Sn impurities can pin the charge order fluctuations, in the sense that finite-frequency fluctuation effects are pushed to lower frequencies and even to static by the quenched disorder\cite{kivelson2003detect, brown2005surface, nie2015fluctuating, brock1994detailed}. For example, Fukuyama, Lee and Rice (FLR) proposed a model to describe the disorder pinning of the phase of CDW\cite{fukuyama1978dynamics, lee2018electric}. In model CDW systems, such as transition metal chalcogenides, this disorder pinning of the phase and amplitude of CDW even above $T$$_{\text{CDW}}$ has been experimentally observed by X-ray photon correlation spectroscopy\cite{sutter2025metastable, yue2020distinction} and  STM\cite{arguello2014visualizing, okamoto2015experimental, dai1993charge} up to 3$T$$_{\text{CDW}}$\cite{arguello2014visualizing}. Moreover, recent studies in Sn-doped CsV$_3$Sb$_5$ support the picture of disorder pinning and forming static short-range CDW.  Nuclear quadrupole resonance (NQR) is a local probe on the bulk and is sensitive to static orderings, and NQR detects the CDW correlations in CsV$_3$Sb$_{5-x}$Sn$_x$\cite{nikolov2025observation}. Therein, static short-range CDW is observed well above $T$$_{\text{CDW}}$ of undoped CsV$_3$Sb$_5$ up to the highest doping level in their experiment ($x = 0.35$), a hallmark of pinning by disorder. This disorder-pinning picture also naturally explains the quasi-1D short-range CDW in Sn-doped CsV$_3$Sb$_5$ by Scanning Tunneling Microscopy and Nuclear Magnetic Resonance\cite{huai2025electronic}, both methods are static probes and thus cannot detect any structure associated with fluctuating CDW. Thus, we believe the picture of disorder pinning and forming static short-range CDW explains our observation of robust CDW correlations beyond the phase boundary of long-range CDW orders in Sn-doped CsV$_3$Sb$_5$. Random defects may favor different local phases of CDW order parameter and in-plane anisotropies, causing the three-state short-range domains revealed by our X-ray diffraction. Since it's not the long-range CDW phase, it may not compete directly with superconductivity, and thus the onset temperature of short-range CDW need not anti-correlate with $T_c$. Finally, the FLR model predicts the total energy gain due to CDW pinning by quenched disorder increases with disorders density\cite{fukuyama1978dynamics, lee2018electric, dai1993charge, gruner2018density}. This explains why the highest temperature where CDW correlations can be detected rises from $x = 0$ to $x = 0.32$ samples, before excessive disorder disrupts any intrinsic correlations and causes this temperature to drop in the $x = 0.68$ sample. 

% In conclusion, we utilize coherent phonon spectrum from TR-reflectivity measurements to study the evolution of CDW in Sn-doped CsV$_3$Sb$_5$. We evidence signatures of CDW correlations far from the temperature and doping level limit of long-range CDW, which can be explained by the quenched disorder introduced by Sn doping that may pin the CDW fluctuations and form static short-range CDW. Remarkably, we confirm CDW correlations up to room temperature in the $x = 0.32$ Sn-doped sample. Our results thus illustrate the important role of doping defects on electronic correlations in Sn-doped CsV$_3$Sb$_5$, and motivates further theoretical and local probe studies into how different dopants intertwine with the complex electronic landscape in Kagome superconductors AV$_3$Sb$_5$. 

\section{IV. Conclusion}

In conclusion, coherent phonon spectra from TR-reflectivity evidence robust CDW correlations persisting far beyond the temperature and doping limit of long-range CDW in Sn-doped CsV$_3$Sb$_5$. These CDW correlations, observed up to room temperature in the $x = 0.32$ sample, can be explained by the quenched disorder introduced by Sn doping that may pin the CDW fluctuations and form static short-range CDW. Our findings highlight the important role of dopant-induced disorder on electronic correlations in Sn-doped CsV$_3$Sb$_5$, and motivates further theoretical and local probe studies into how different dopants intertwine with the complex electronic landscape in Kagome superconductors $A$V$_3$Sb$_5$.

\section*{Acknowledgment}
L.W. acknowledges support from the Department of Energy Early Career Research Program Award under No. DE-SC0026208. The construction of the pump-probe setup was supported by the Air Force Office of Scientific Research under award no. FA9550-22-1-0410. Q.D. was mainly supported by the Vagelos Institute of Energy Science and Technology graduate fellowship, Dissertation Completion Fellowship and also partly supported by the National Science Foundation EPM program under grant no. DMR-2213891 (ended in 2025). S.D.W. and A.C.S. gratefully acknowledge support via the UC Santa Barbara NSF Quantum Foundry funded via the Q-AMASE-i program under award DMR-1906325. This work is based on research conducted at the Center for High-Energy X-ray Sciences (CHEXS), which is supported by the National Science Foundation (BIO, ENG and MPS Directorates) under award DMR-2342336.

\bibliography{Sn-doped_CsV3Sb5_TRR_refs}

\clearpage

\end{document}